\documentstyle[epsfig]{aipproc}
\def\Journal#1#2#3#4{{\em #1}{\bf #2}, #3 (#4)}
\def\PTP{Prog.~Theor.~Phys.~}

\def\NPA{Nucl.~Phys. \bf{A}}
\def\NPB{Nucl.~Phys. \bf{B}}
\def\PLB{Phys.~Lett. \bf{B}}

\def\PRL{Phys.~Rev.~Lett.~}
\def\PRD{Phys.~Rev. \bf{D}}
\def\ZPC{Z.~Phys. \bf{C}}

\def\PR{Phys.~Rev.~}

\begin{document}
\title{On existence of the $\sigma$(600)\\
{\large\bf Its physical implications and related problems}}

\author{Shin Ishida}
\address{Atomic Energy Research Institute,
College of Science and Technology\\
Nihon University, 
Kanda-Surugadai, Chiyoda, Tokyo 101, Japan}

\maketitle

\begin{abstract}
We make a re-analysis of I=0 $\pi\pi$
scattering phase shift $\delta_0^0$
through a new method of $S$-matrix parametrization
(IA; interfering amplitude method),
and show a result suggesting strongly for 
the existence of $\sigma$-particle -- long-sought 
Chiral partner of $\pi$-meson.
Furthermore, through the phenomenological 
analyses of typical production processes
of the $2\pi$-system, the $pp$-central collision      
and the $J/\Psi\rightarrow\omega\pi\pi$ decay,
by applying an intuitive formula as sum of Breit-Wigner amplitudes,
(VMW; variant mass and width method),
the other evidences for the $\sigma$-existence are given.

The validity of the methods used in the above analyses is  
investigated, using a simple field theoretical model,
from the general viewpoint of unitarity and the applicability of
final state interaction(FSI-) theorem, especially
in relation to the ``universality'' argument.
It is shown that the IA and VMW are 
obtained as the physical state representations of
scattering and production amplitudes, respectively.
The VMW is shown to be an effective
method to obtain the resonance properties from production processes, 
which generally have the unknown strong-phases.
The conventional analyses based on the ``universality''
seem to be powerless for this purpose.\footnote{
This is a brief review of
 essential points given in the
 four contributions in parallel sessions;
(i)``re-analysis of $\pi\pi$/K$\pi$ phase shift..'' by T.~Ishida,
(ii) ``existence of $\sigma$/$\kappa$ particle..'' by M.Y.~Ishida,
(iii) ``$\sigma$ particle in production..'' by K.~Takamatsu, and
(iv) ``relation between scattering and..'' by M.Y~Ishida.
These are referred as [{\bf psa}], [{\bf lsm}], [{\bf prd}], 
and [{\bf rel}], respectively.
}
\end{abstract}

\noindent \underline{\large\bf Introduction}

The light iso-singlet scalar $\sigma$-meson
is a very important particle which is predicted to exist 
as a chiral partner of the Nambu-Goldstone $\pi$-meson,
corresponding to the dynamical breaking of chiral symmetry
existing in the massless limit of QCD\cite{rf:NJL},
with  mass$\approx 2m_q$ ($m_q$ being
the constituent quark mass).
This $\sigma$ gives quarks constituent masses,
and in this sense $\sigma$ may be called ``Higgs particle
in QCD." The possible existence of 
the $\sigma$-particle has been suggested from various viewpoints
both theoretically and phenomenologically.
In particular the importance of $\sigma$ 
in relation with the D$\chi$SB has been
argued extensively by Refs.\cite{rf:sca,rf:kunih0}

However, the existence of $\sigma$ as a resonant particle 
has not yet been generally accepted.
A major reason for this is due to the analyses
of $\pi\pi$ phase-shift obtained from 
CERN-M\"unich experiment\cite{rf:CM} in 1974,
in which the I=0 $\pi\pi$ $S$-wave phase shift $\delta_0^0$ up to
$m_{\pi\pi}=1300$ MeV turned out to be only $270^\circ$.
After subtracting a rapid contribution of 
the resonance $f_0(980)(180^\circ )$,
there remains only $90^\circ$, being insufficient
for $\sigma$ around $m=2m_q=500\sim 600$ MeV.
Most analyses thus far made on it
have yielded conclusions against the existence 
of $\sigma$\cite{rf:morg}.
As a result the light $\sigma$-particle had been disappeared from
the list of PDG since the 1976 edition\cite{rf:pdg76}.

On the other hand, in the recent
$pp$-central collision experiment,
a huge event concentration in I=0 $S$-wave $\pi\pi$-channel
is seen\cite{rf:GAMS} 
in the region of $m_{\pi\pi}$ around 500$\sim$600 MeV,
which is too large to be explained as a simple
``background'' and seems strongly suggest the existence of $\sigma$.
Actually it is shown that the characteristic shape 
of $\pi^0\pi^0$ effective mass spectra below 1 GeV is
able to be explained\cite{rf:taku} by the two, 
$\sigma$- and $f_0$-, Breit-Wigner resonant amplitudes
interferring mutually.

\begin{figure}[t]
\epsfysize=12.5 cm
\centerline{\epsffile{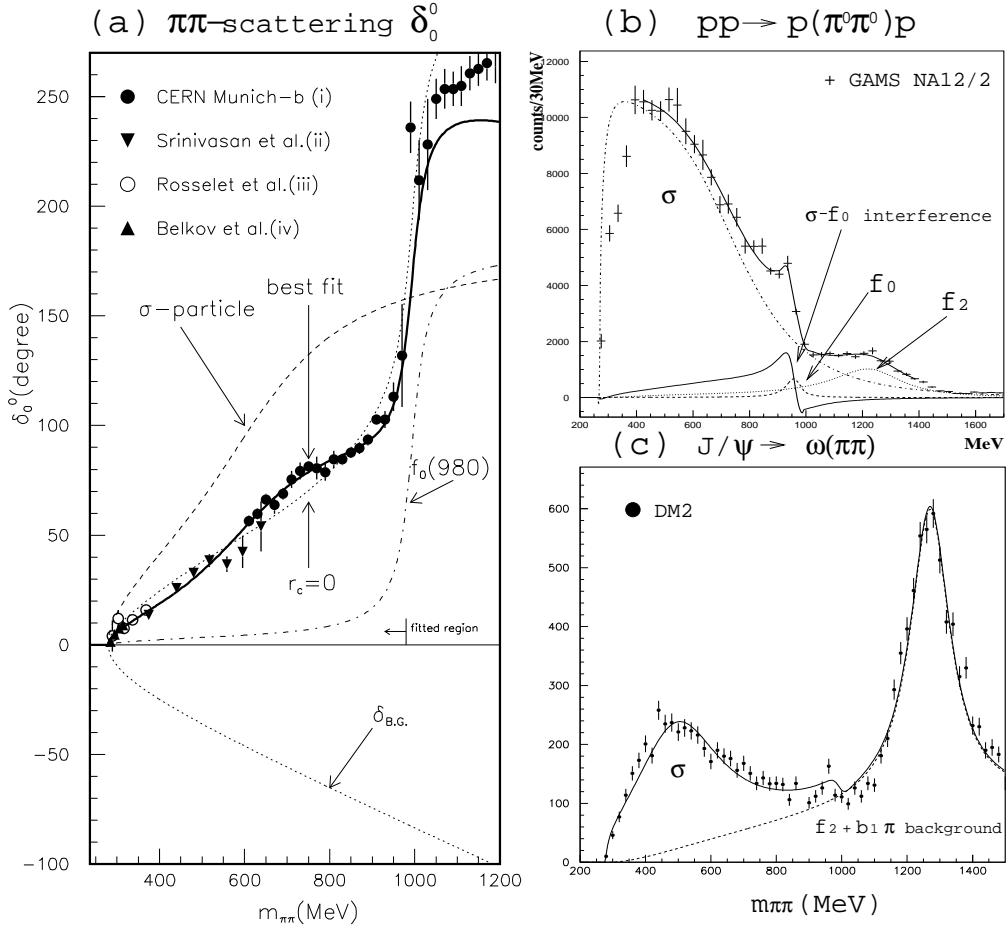}}
\caption{Phenomenological observation of
$\sigma$-particle both in scattering and production processes.
(a) $I$=0 $\pi\pi$ scattering phase shift,
(b) $\pi^0\pi^0$-effective mass distribution in $pp$-central collision
(GAMS NA12/2), (c) $\pi\pi$-effective mass distribution in 
$J/\psi\rightarrow\omega\pi\pi$ decay (DM2).
}
\label{fig:pob}
\end{figure}
However the claim, $\sigma$-existence in the production
processes, had been criticized from the so-called ``universality"
argument\cite{rf:pennington}: ``Unitarity requires a
resonance that decays to $\pi\pi$, for example,
has to couple in the same way to this final state
whether produced in $\pi\pi$ scattering or centrally
in $pp\rightarrow pp(\pi\pi )$." ``Thus claims of a 
narrow $\sigma$(500) in the GAMS results cannot be
correct as no such state is seen in $\pi\pi$ scattering."\\


\noindent \underline{\large\bf 
Phenomenological Observation of $\sigma$-particle}

On the contrary, 
we have recently made 
a re-analysis\cite{rf:pipip,rf:further} of 
the $\pi\pi$ phase shift and found a strong evidence 
for existence of $\sigma$-particle[{\bf psa}].\footnote{
Independently the other several groups have performed
the re-analyses of the phase shift, leading also to positive 
conclusion\cite{rf:achasov,rf:kamin,rf:torn,rf:hara} 
for $\sigma$-existence.
Reflecting this situation 
the $\sigma$-particle has been revived in the list of
latest edition of PDG,\cite{rf:pdg96}
after missing for twenty years,
with  somewhat a hesitating label 
``$f_0(400\sim 1200)$ or $\sigma$". 
}
\begin{table}[b]
\caption{
Comparison between the fit with $r_c\neq 0$
and with $r_c= 0$ in our PSA. The latter corresponds
to the conventional analyses thus far made.
}
\begin{center}
\begin{tabular}{|c|c||c|}
\hline
\hline
 & $r_c\neq 0(\chi^2/N_f=23.6/30)$
               & $r_c= 0(\chi^2/N_f=163.4/31)$    \\
\hline
    & $\delta^{\rm tot}=\delta_{f_0(980)}
      +[\delta_{\sigma (600)}+\delta_{\rm BG}
        ]^{\rm pos.}$
    & $\delta^{\rm tot}=\delta_{f_0(980)}
      +[\delta_{\rm BG}^{\rm pos.}]$ \\
    & $\sigma (600)$ 
     &  ``$\sigma$" (equivalent to $\epsilon (900)$\cite{rf:morg})\\
\hline 
$m_\sigma$ & $585\pm 20({\tiny 535\sim 675})$  &  920\\ 
$\Gamma_\sigma^{(p)}$ & $385\pm 70$ & 660\\
$\sqrt{s_{\rm pole}}/$MeV & $(602\pm26)-i(196\pm 27)$ 
                    & 970-i320  \\
$r_c$   &  3.03$\pm$0.35 GeV$^{-1}$  &  -- \\
   ~    &  (0.60$\pm$0.07 fm)        & (--)\\
\hline
\end{tabular}
\end{center}
\label{tab:conventional}
\end{table}
The reasons which led us to a different result from
the conventional one\cite{rf:morg}, 
even with the use of the same data 
of phase shifts, are twofold: On one hand technically,
we have applied a new method of Interferring Breit-Wigner 
Amplitude (IA-method) for the analyses,
where the ${\cal T}$-matrix (instead of ${\cal K}$-matrix 
in the conventional treatment) for multiple resonance case
is directly represented by the respective Breit-Wigner
amplitudes in conformity with unitarity,
thus parametrizing the phase shifts directly in terms of
physical quantities, such as masses and coupling constants 
of the relevant resonant particles.
On the other hand physically, we have introduced
a ``negative background phase" $\delta_{\rm BG}$
of hard core type, $\delta_{\rm BG}(s)$=$-|{\bf p}_1|r_c $,
whose possible origin is discussed later.
The results of our re-analyses are given in Fig.~\ref{fig:pob}(a), while
in Table~\ref{tab:conventional} I compare
the essential points and the results of our analysis with those of 
the conventional one. 
In our analysis the introduction
of repulsive $\delta_{\rm BG}$ with $r_c\sim 3$GeV${}^{-1}$
(0.60fm, about the structural size of pion)
plays a crucial role for the existence of $\sigma$(600).
The sum of the large attractive $\delta_{\sigma (600)}$, 
contribution due to $\sigma$(600),
and the large repulsive $\delta_{\rm BG}$ gives a small
positive phase shift,
which was treated,
in the conventional analysis,\cite{rf:morg}
as a background (or broad $\epsilon ($900))
contribution $[\delta_{\rm BG}^{\rm pos.}]$.
Note that the fit with $r_c$=0 in our analyses corresponds to
the conventional analyses without the repulsive
$\delta_{\rm BG}$ thus far made.
In this case the mass and width of ``$\sigma$"
becomes large, and the ``$\sigma$"-Breit-Wigner
formula can be regarded as an  
effective range formula 
describing a positive background phase.
The corresponding pole position
is close to that of $\epsilon$(900) in Ref.~\cite{rf:morg}.
In this case the bump-like structure 
around $500\sim 600$ MeV in $\delta_0^0$ 
cannot be reproduced, as shown in Fig.~\ref{fig:pob}(a), and
gives $\chi^2$ = 163.4, worse by 140 than the best fit.
This seems strongly to suggest the $\sigma$-existence.

\begin{table}[t]
\caption{Observed mass and width 
of $\sigma$-particle in production processes.} 
\begin{center}
\begin{tabular}{|l|c|c|c|}
\hline
~&mass(MeV)&$\Gamma_{\pi\pi}^p$(MeV) \\
\hline
$pp$-central (GAMS NA12/2) & 590$\pm$10 & 710$\pm$30 \\
$J/\psi$ decay (DM2)       & 480($\pm$5$_{st}$) & 325($\pm$10$_{st}$) \\
\hline
\end{tabular}
\end{center}
\label{tab:obs}
\end{table}
We also present[{\bf prd}] possible evidences
for the existence of the $\sigma$ particle 
as an intermediate state of 
the $\pi\pi$ system in production processes,
by analyzing the data obtained through the 
{\em pp} central collision experiment by
GAMS\cite{rf:GAMS,rf:taku} and 
the data in the 
$J/\psi\rightarrow\omega\pi\pi$ decay 
reported by DM2 collabration\cite{rf:JPsi}.
As is seen in Fig.~\ref{fig:pob}(b) and (c),
in each process there is a huge concentration of events in
the $\pi\pi$ effective mass spectrum below 1 GeV,
which is able to be simply explained as the $\sigma$-resonance. 
In the analyses we apply the 
Variant Mass and Width(VMW)-method,\footnote{
It was named\cite{rf:sawa} historically after the following reason.
The mass and width of ``a" resonant particle,
which is misinterpreted as one resonance,
instead of actual two overlapping resonances, 
are observed variantly depending upon the respective processes.
} 
where the production amplitude is represented by a 
sum of the $\sigma$ and $f_0$ Breit-Wigner amplitudes 
with relative phase factors
\begin{eqnarray}
\frac{r_\sigma e^{i\theta_\sigma}}{m_\sigma^2-s-i\sqrt{s}\Gamma_\sigma (s)}
+\frac{r_f e^{i\theta_f}}{m_f^2-s-i\sqrt{s}\Gamma_f (s)},
\label{eq:VMW}
\end{eqnarray}
It is notable that, with this rather simple formalizm, 
not only the effective mass distribution,
but also the characteristic angular distribution are excellently
reproduced. It seems to us that $\sigma$-particle is
observed as a huge event concentration of $S$-wave states
in these production processes. 
The obtained mass and width of $\sigma$,
with those of $f_0$ being fixed, 
are given in Table.\ref{tab:obs}.\\


\noindent \underline{\large\bf
Relation between Scattering and Production Amplitudes}

In treating the $\pi\pi$-scattering and production amplitudes,
there are two general problems to be taken into account:
The scattering amplitude ${\cal T}$ must satisfy
the unitarity
${\cal T}-{\cal T}^\dagger = 2i{\cal T}\rho{\cal T}^\dagger$,
and the production amplitude ${\cal F}$ must have, 
in case that the initial state has no strong phase,
 the same 
phase as ${\cal T}$:
${\cal T}\propto e^{i\delta}\rightarrow
{\cal F}\propto e^{i\delta}$
(FSI; Final-State-Interaction
theorem\cite{rf:watson}).
Conventionally, the more restrictive relation
between  ${\cal F}$ and  ${\cal T}$
is required on the basis of the 
``universality,"\cite{rf:morg,rf:pennington}
\begin{eqnarray}
{\cal F} &=& \alpha (s){\cal T}
\label{eq:FaT}
\end{eqnarray}
with a slowly varying real function $\alpha (s)$ of $s$.
The criticism on our results of phenomenological
analysis on the production process using the 
VMW-method was raised\cite{rf:pennington}
along this line, as was mentioned.
I have already shown, through the reanalysis
of the $\pi\pi$-scattering, 
that there exists surely the $\sigma$-pole 
in ${\cal T}$. Accordingly a main reason of the above 
criticism loses its reason.
However, there has been remained a problem
in the VMW-method applied to ${\cal F}$;
whether it is consistent to the FSI-theorem, or not.

In the following I re-examine the relation
between ${\cal F}$ and ${\cal T}$ 
concretely, by using a simple model\cite{rf:aitchson,rf:chung}.
In the NJL-type model as a low energy effective theory of QCD,
(and in the linear sigma model, L$\sigma$M, obtained as its local limit),
or in the constituent quark model, 
the pion $\pi$ and
the resonant particles such as $\sigma$(600) or $f_0(980)$
are the color-singlet $q\bar q$-bound states and  
are treated equally. These  
``intrinsic quark dynamics states,"
denoted as $\bar \pi,\ \bar \sigma ,\  \bar f$
are stable particles with zero widths and appear from the beginning. 
Actually these particles 
have structures and interact with one another
(and a production channel ``$P$'') 
through the residual strong interaction:
\begin{eqnarray}
{\cal L}^{\rm scatt}_{\rm int} &=&
\sum \bar g_\alpha\bar\alpha\pi\pi
+\bar g_{2\pi}(\pi )^4\ \ \ \  
({\cal L}^{\rm prod}_{\rm int}=\sum\bar\xi_\alpha\bar\alpha ``P"
+\bar\xi_{2\pi}\pi\pi  ``P") .
\label{eq:Lint}
\end{eqnarray}
Due to this, these bare states change\cite{rf:achasov} into
the physical states, denoted as
$\pi (=\bar\pi ),\ \sigma$ and $f$ with finite widths.
In the following we consider only the
virtual two-$\pi$ meson effects 
for the resonant $\sigma$ and $f$ particles.



There are following 3-ways representing scattering amplitudes,
corresponding to the three sorts of the basic 
states for describing the resonant particles:

\begin{description}

\item[1.\ ] {\it Intrinsic\ quark-dynamics\ states\ (bare states)
representation}\\
In the bases of zero-width bare states,
denoted as $|\bar \alpha\rangle$,
the $\pi\pi$-scattering amplitude ${\cal T}$ 
 is represented in terms of 
the $\pi\pi$-coupling constants $\bar g_{\bar\alpha}$
 and the propagator matrix $\bar\Delta$ as
\begin{eqnarray}
   {\cal T} &=& \bar g_{\bar\alpha}\bar\Delta_{\bar\alpha\bar\beta}
   \bar g_{\bar\beta};\ \ \ 
  \bar\Delta_{\bar\alpha\bar\beta}^{-1}
  = (\bar M^2-s-i\bar G)_{\bar\alpha\bar\beta}.
   \label{eq:TinB}
\end{eqnarray}
The real and imaginary parts of the squared mass matrix
take the non-diagonal forms, which mean
the bare states have indefinite masses and life times.
The imaginary part of the inverse propagator is 
$   \bar G_{\bar\alpha\bar\beta}
     = \bar g_{\bar\alpha}\rho\bar g_{\bar\beta}\ 
(\rho =\sqrt{1-4m_\pi^2/s}/16\pi $ being the state density),
then our ${\cal T}$ is easily shown to satisfy 
the unitarity.

\item[2.\ ] ``${\cal K}$-{\it matrix"\ states representation},\\
The real part of the $\bar\Delta^{-1}$ is symmetric and
can be diagonalized by an orthogonal transformation:
It transforms the bare states 
$|\bar\alpha\rangle$ into the
``${\cal K}$-matrix" states 
 as $|\tilde \alpha\rangle\equiv |\bar\alpha\rangle 
                               o_{\bar\alpha\tilde\alpha}$.
Correspondingly the
${\cal T}$ is represented by
\begin{eqnarray}
{\cal T} &=& \tilde g_{\tilde\alpha}
\tilde\Delta_{\tilde\alpha\tilde\beta}
\tilde g_{\tilde\beta};\ 
\tilde\Delta_{\tilde\alpha\tilde\beta}^{-1}
=(\Delta^{-1}_{{\cal K}}-i\tilde G)_{\tilde\alpha\tilde\beta},
{\Delta^{-1}_{{\cal K}}}_{\tilde\alpha\tilde\beta}
= (\tilde m^2_{\tilde\alpha}-s)
\delta_{\tilde\alpha\tilde\beta},
\tilde G_{\tilde\alpha\tilde\beta}=\tilde g_{\tilde\alpha}
\rho\tilde g_{\tilde\beta},
\label{eq:TinK}
\end{eqnarray}
where the coupling constant
 $\tilde g_{\tilde\alpha}
(=\bar g_{\bar\alpha}o_{\bar\alpha\tilde\alpha})$
is real.
These states have definite masses, but indefinite life times.
The propagator $\tilde\Delta$ is able to be expressed in the form
representing concretely 
the repetition of the $\pi\pi$-loop, 
as 
\begin{eqnarray}
\tilde\Delta =(1-i\Delta_{{\cal K}}\tilde G
)^{-1}\Delta_{{\cal K}}
&=& \Delta_{{\cal K}}+i\tilde\Delta\tilde G
\Delta_{{\cal K}}.
\label{eq:DinK2}
\end{eqnarray}
Then the ${\cal T}$ takes the same form\footnote{
From the viewpoint of the present 
field-theoretical model, 
this ``${\cal K}$-matrix,"
Eq.(\ref{eq:Krep}), has a physical meaning
as the  propagators of bare particles
with infinitesimal imaginary widths,
$\tilde m_{\tilde\alpha}^2\rightarrow
\tilde m_{\tilde\alpha}^2-i\epsilon$,
while the original 
${\cal K}$-matrix in potential theory is purely real
and has no direct meaning.
 }
 as the ${\cal K}$-matrix in potential theory:
\begin{eqnarray}
{\cal T} = {\cal K}+i{\cal T}\rho{\cal K}
= {\cal K}(1-i\rho{\cal K})^{-1};\ \ \
 {\cal K} = \tilde g_{\tilde\alpha}
\Delta_{{\cal K}\tilde\alpha\tilde\beta}
\tilde g_{\tilde\beta}
=\tilde g_{\tilde\alpha}
(\tilde m_{\tilde\alpha}^2-s)^{-1}
\tilde g_{\tilde\alpha}.
\label{eq:Krep}
\end{eqnarray}

\item[3.\ ] {\it Physical\ resonant\ states representation},\\
The imaginary part of 
$\tilde\Delta^{-1}$
in the ${\cal K}$-matrix state representation,
which was remained in a non-diagonal form,
can be diagonalized by 
a complex orthogonal\cite{rf:aitchson,rf:rosen} matrix $u$,
satisfying ${}^tuu=1$.
It transforms 
$|\tilde\alpha\rangle$
into the unstable physical states as $|\alpha\rangle\equiv 
|\tilde\alpha\rangle u_{\tilde\alpha\alpha}$.
It is to be noted that
the transformation is not unitary and 
$\langle\alpha |\neq (|\alpha\rangle )^\dagger$.
Correspondingly 
the ${\cal T}$-matrix is represented by
\begin{eqnarray}
{\cal T} &=& F_{\alpha}\Delta_{\alpha\beta}F_{\beta}
=\sum_\alpha F_\alpha (\lambda_\alpha-s)^{-1}F_\alpha ;\ \ \ 
F_{\alpha}(\equiv \tilde g_{\tilde\beta}u_{\tilde\beta\alpha})
\label{eq:TinP}
\end{eqnarray}
where the $\lambda_\alpha$ is the physical squared mass of the
$\alpha$-state, and  the 
$F_{\alpha}$
are the physical coupling constants, which are generally complex.
The physical state has a 
definite mass and life time,
and be observed as a resonant particle
directly in experiments.

\end{description}

In the following 
I show how the formulas
in the IA and VMW methods 
satisfying FSI theorem[{\bf rel}] are derived effectively 
in the physical state representation. 
We start from the ``${\cal K}$-matrix" states,
which are able to be identified
with the bare states
$|\bar \alpha\rangle (\equiv |\tilde \alpha\rangle )$
without loss of essential points, 
since the reality
of the coupling constant is unchanged
through the orthogonal transformation.
The real part of the mass correction generally does not
have sharp $s$-dependence, and 
the $\tilde g$ is 
almost $s$-independent, except for 
the threshold region. Then,  
in the two( $\bar\sigma ,\ \bar f$) 
resonance-dominative case,
the scattering amplitude ${\cal T}$ 
is given by Eq.(\ref{eq:Krep}) as
\begin{eqnarray}
{\cal T}
  &=& 
{\cal K}/(1-i\rho{\cal K});\ \ \ \ 
{\cal K}
=\bar g_{\bar\sigma}^2/(\bar m_{\bar\sigma}^2-s)
+\bar g_{\bar f}^2/(\bar m_{\bar f}^2-s).
\label{eq:Kresrep}
\end{eqnarray}
The production amplitude ${\cal F}$ is obtained,
by replacing the one of respective
scattering-coupling-constant $\bar g$
with the production coupling-constant $\bar\xi$, as
\begin{eqnarray}
{\cal F} &=& {\cal P}/(1-i\rho{\cal K});\ \ \ \ 
{\cal P}=\bar\xi_{\bar\sigma}\bar g_{\bar\sigma}/(
\bar m_{\bar\sigma}^2-s)
+\bar\xi_{\bar f}\bar g_{\bar f}
/(\bar m_{\bar f}^2-s).
\label{eq:Presrep}
\end{eqnarray}
The FSI-theorem is automatically satisfied
since both ${\cal K}$ and ${\cal P}$
can be treated as real and
the phases of ${\cal T}$ and ${\cal F}$
come from the common factor $(1-i\rho{\cal K})^{-1}$.

In the physical state representation 
 the  ${\cal T}$ is given by 
\begin{eqnarray}
{\cal T} = \frac{F_\sigma^2}{\lambda_\sigma-s}  
+\frac{F_f^2}{\lambda_f-s} = \frac{g_\sigma^2}{\lambda_\sigma-s}  
+\frac{g_f^2}{\lambda_f-s}+2i\rho 
\frac{g_\sigma^2}{\lambda_\sigma-s}\frac{g_f^2}{\lambda_f-s}.
\label{eq:TresinP}
\end{eqnarray}
This is just the form of
scattering amplitude, applied in IA-method[{\bf psa}].
The $\lambda_\alpha$($\alpha =f,\sigma$)
 is identified to $M_\alpha^2-i\rho g_\alpha^2$ 
appearing in usual Breit-Wigner formula.
Thus we define the physical mass $M_\alpha$ 
and the real physical coupling $g_\alpha$ 
($g_\alpha^2\equiv -{\rm Im}\ \lambda_\alpha /\rho$).
 
Similarly 
in the physical state representation the ${\cal F}$ 
is given by
\begin{eqnarray}
{\cal F} &=& 
\frac{r_\sigma e^{i\theta_\sigma}}{\lambda_\sigma-s}  
+\frac{r_f e^{i\theta_f}}{\lambda_f-s}.
\label{eq:VMWderive}
\end{eqnarray}
The $r_\alpha$ and $\theta_\alpha$ are 
expressed in terms of $\bar g_{\bar\alpha},\ \bar \xi_{\bar\alpha},$ 
and $\lambda_\alpha$ and shown to be almost 
$s$-independent except for the threshold region.
Thus the Eq.(\ref{eq:VMWderive}) has the same form  
as Eq.(\ref{eq:VMW}) applied in VMW-method.
However, we must note on the following:\ \ 
In the VMW-method essentially the 
three new parameters,
$r_\sigma ,\ r_f$ and the relative phase 
$\theta (\equiv\theta_\sigma -\theta_f)$,
independent of the 
scattering process,
characterize the relevant production processes.
Presently they
are represented by the 
two production coupling constants, $\bar\xi_{\bar\sigma}$
and $\bar\xi_{\bar f}$.
Thus, among the three parameters in VMW-method there exists
one constraint due to the FSI-theorem.
The corresponding considerations 
in the case with the non-resonant background phase 
are also given in [{\bf rel}].

Here it should be noted that 
the FSI-theorem is only applicable
to the case of the initial state having no strong phase.
This type of initial strong phases 
generally exists in all processes 
under the effect of strong interactions, which is effectively 
able to be introduced in the VMW-method by
substitution of 
$
\bar r_{\bar\alpha} \rightarrow  \bar r_{\bar\alpha}
e^{i\bar\theta_{\bar\alpha}^{\rm strong}}.
$
However, we have few knowledge on the initial 
phases, and we are forced to treat the parameters 
in VMW-method as being effectively free.

\begin{figure}[t]
 \epsfysize=12. cm
 \centerline{\epsffile{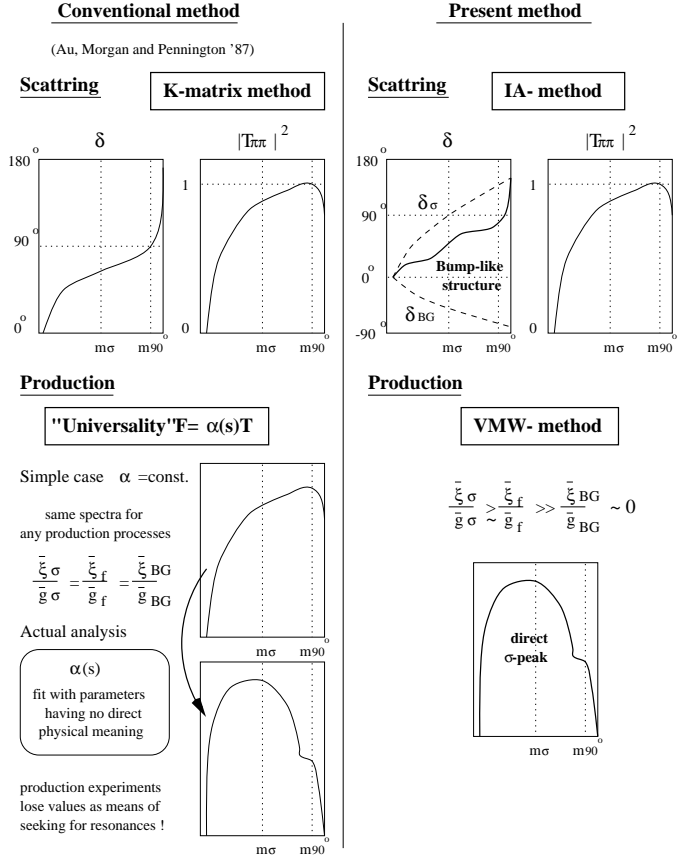}}
 \caption{Analyses by IA- and  VMW-methods
compared with the conventional analyses based on 
the ``Universality" of $\pi\pi$-scattering}
\label{fig:compare}
\end{figure}
The way of our 
analyses of scattering and production processes
are compared with that of 
the conventional analyses based on 
the ``Universality" argument 
pictorially in Fig. \ref{fig:compare}. 
The $\pi\pi$-scattering is largely affected by
the effect of the non-resonant repulsive background, 
and the  ${\cal T}$ cannot be described only by 
usual Breit-Wigner amplitudes
with a non-derivative coupling.
The spectrum of ${\cal T}$ shows a very wide peak
around the $\sqrt{s}\simeq 850$ MeV, at which the phase
 $\delta_0^0$ passes through 90 degrees,
and then falls down rapidly.
In contrast 
the spectra of ${\cal F}$
in the $pp$-central collision and
the $J/\Psi\rightarrow\omega\pi\pi$-decay 
have peaks
at around $\sqrt{s}= 
m_\sigma (500\sim 600$MeV).
In the conventional way, with the universality 
relation ${\cal F}=\alpha {\cal T},$
the ${\cal T}$ is first analyzed and the 
phase shift $\delta$ around $\sqrt{s}=m_\sigma$
is interpreted as due to, instead of $\sigma$-contribution,
the background. 
Then ${\cal F}$ is analyzed with
the $\alpha (s)$ arbitrarily chosen 
with the polynomial form,
$\alpha (s) = \sum \alpha_ns^n$.
In the most simple
case with $\alpha =$const. the universality relation
implies that
$\bar \xi_\sigma=\alpha \bar g_\sigma$,
$\bar \xi_f=\alpha \bar g_f$ and 
$\bar \xi_{2\pi}=\alpha \bar g_{2\pi}$,
that is, all the production couplings
to be 
proportional to the corresponding $\pi\pi$-couplings,
and the spectra of ${\cal F}$ and ${\cal T}$
becomes the same.
Actually they are different
and the difference is fitted by the 
$\alpha_n$.
The masses and widths of resonances  
 are determined only from the 
$\pi\pi$-scattering, and 
the analyses of ${\cal F}$ on any production
process become 
 nothing but the determination of the
$\alpha_n$ for respective processes,
 which have no direct physical meaning. 
Thus all the production experiments 
lose their values in seeking for new resonances. 
On the other hand, in the VMW-method,
only the physically meaningful parameters are introduced.
The $\bar \xi_{\bar\sigma}$, $\bar \xi_{\bar f}$
and $\bar \xi_{2\pi}(s)$ are independent
parameters of the $\pi\pi$-scattering,
and the difference between the spectra of
${\cal F}$ and ${\cal T}$ is explained
intuitively by supposing the relations among the   
coupling constants such as
$
\bar\xi_{\bar\sigma}/\bar g_{\bar\sigma}  \gg  
\bar\xi_{2\pi}/\bar g_{2\pi},
$
that is, 
the ratio of background effects to the $\sigma$-effects 
are weaker in 
the production processes
than in the scattering process.
Thus in this case  
the large low-energy peak structure in $|{\cal F}|^2$ 
shows directly the $\sigma$-existence.
In this situation
the properties of $\sigma$
can be obtained more precisely 
in production processes
than in scattering processes.
Here I should like to note that 
this difference between two methods may reflect their
basic standpoints: In the ``universality" argument
only the stable (pion) state consists in 
the complete set of meson
states, while
the $\bar \sigma$ and $\bar f$,
in addition to pion,
are necessary as  bases of the complete set in VMW-method.\\


\noindent \underline{\large\bf Physics connected with $\sigma$-existence}

\paragraph*{{\large Origin of repulsive core}\ \ }

In our phase shift analyses the repulsive 
$\delta_{\rm BG}$ of hard core type 
 introduced phenomenologically 
plays an essential role.
This type of $\delta_{\rm BG}$
was also reported historically
in the $\alpha\alpha$-scattering\cite{rf:acore}
and in the $NN$-scattering\cite{rf:Ncore},
whose origin may be related to 
the Fermi-statistics property 
of respective constituent particles. 

The repulsive core in the $\pi\pi$- and $K\pi$-scatterings
seems to have a strong connection to the $\lambda\phi^4$-interaction
in L$\sigma$M:
It represents a contact zero-range interaction
and is strongly repulsive,
and has plausible properties
as the origin of repulsive core.
Moreover, the magnitudes of core radii are shown to be 
explained qualitatively by this term[{\bf lsm}].
In NJL-type model the composite pion field $\phi (x)$ is defined
in the ``local limit'' 
from the constituent quark field  $q(x)$ by
$\phi (x)=\bar q(x)i\gamma_5{  \tau}q(x)$,
and the $\lambda\phi^4$-interaction is obtained from
the quark box-diagram.
The repulsiveness of the 
$\lambda\phi^4$-interaction, $\lambda >0$, 
is explained by the loop factor -1
due to the Fermi-statistics property of
constituent quarks.

\paragraph*{{\large $q\overline{q}$-meson spectra and ``Chiralons''}\ \ }

Taking the $SU(3)$ flavor symmetry into account, 
it is now natural to expect the existence of a 
scalar meson nonet.
We also analyze[{\bf psa}] the {\em I}=1/2 $K\pi$-scattering 
phase shift from a similar 
standpoint to the $\pi\pi$ system, 
and actually show\cite{rf:piK} that its behavior 
is consistent with the existence of an
{\em I}=1/2 scalar meson, the $\kappa$(900) meson.
The $\sigma$(600) and $\kappa$(900), and the 
observed resonances $a_0(980)$ and $f_0(980)$, are shown[{\bf lsm}] to 
have almost plausible properties 
as the members of the $\sigma$-nonet
in the SU(3)L$\sigma$M and 
the SU(3)L$\sigma$M with $\rho$- and $a_1$-nonets.
The $\sigma$-nonet 
forms with the pseudoscalar 
$\pi$-nonet the linear representation of 
chiral symmetry.
This result implies that 
the chiral symmetry plays the stronger role
than ever thought 
in understanding the strong interaction, especially 
not only the low energy theorems 
derived through the non-linear realization, 
but also the spectroscopy and reactions related with 
all the mesons with masses 
below and around $\sim$ 1 GeV
through the linear realization.

Here I should like to mention that the $\sigma$-nonet is to be 
discriminated from the ${}^3P_0$ scalar nonet:\ 
There are well-known two contrasting views on $q\bar{q}$-meson 
spectra.
One is based on LS-coupling scheme of non-relativistic
quark model.
In boosted LS-coupling scheme\cite{rf:spin} 
 the covariant $q\bar{q}$-wave function(WF)
is obtained by boosting the non-relativistic space and spin 
WF's, $f({\bf r})\otimes \chi_i\tilde{\chi}^j$
separately into
$f(\tilde{r}_\mu )\otimes U_\alpha^\beta$. 
The Bargmann-Wigner function, $U_\alpha^\beta$, is represented 
by a direct product,
$U_\alpha^\beta =u_\alpha ({\bf v})\bar{v}^\beta ({\bf v})$, 
of free Dirac spinors of quark
and anti-quark
with the same velocity $v_\mu$ as the meson.
The $U$ includes the pseudoscalar and vector component, but no scalar
since $\langle \bar{v}({\bf v}) u({\bf v}) \rangle =0$.
The space-time WF $f$ is a function 
of the  $\tilde{r}_\mu\equiv (\delta_{\mu\nu}+v_\mu v_\nu )r_\nu$ 
( $r_\mu$ and $v_\mu$ being the 
relative space-time coordinate and the velocity of meson, respectively).

The other view is based on chiral symmetry
and has a covariant framework from the beginning. 
In the NJL model the local $\sigma^i$ field is defined by
$\sigma^i(x)\equiv\bar{\psi}(x)\lambda^i\psi (x)$,
missing the freedom of relative space-time coordinate  
$r_\mu$ which is required to describe
orbital excitations. 
The local $\sigma^i(x)$ is only describable, 
so to speak, L=0 states.
Being based on a similar consideration the axial $a_1$-nonet
as a partner of $\rho$-nonet
is also expected to exist besides
of the $^3P_1$-nonet.

\end{document}